\documentclass[10pt, conference, letterpaper]{IEEEtran}

\usepackage{cite}
\usepackage{amsmath,amssymb,amsfonts}
\usepackage{algorithmic}
\usepackage{graphicx}
\usepackage{textcomp}
\usepackage{xcolor}
\usepackage{algorithm}
\usepackage{booktabs}
\usepackage{multirow}
\usepackage{colortbl}
\usepackage{comment}
\usepackage{enumitem}
\usepackage{adjustbox}
\usepackage{tabularx}
\usepackage{array}
\usepackage{placeins}
\usepackage[hidelinks]{hyperref}
\usepackage{orcidlink}

\newtheorem{definition}{Definition}
\newtheorem{example}{Example}

\begin{document}

\title{DepTGL: A Parallel Framework for Memory-based TGNN Training with Adaptive Temporal Data Dependency Management}

\author{
    \IEEEauthorblockN{
        Linfang Chen\,\orcidlink{0009-0005-7730-1498}$^\S$, 
        Zhen Song\,\orcidlink{0009-0004-8737-2727}$^\S$, 
        Lei Liu\,\orcidlink{0000-0002-4646-2054}$^\S$, 
        Yu Gu\,\orcidlink{0000-0001-7422-6254}$^\ddagger$, 
        Yushuai Li\,\orcidlink{0000-0002-3043-3777}$^\dagger$, \\
        Yanfeng Zhang\,\orcidlink{0000-0002-9871-0304}$^\ddagger$, 
        Lizhen Cui\,\orcidlink{0000-0002-8262-8883}$^\S$, 
        Ge Yu\,\orcidlink{0000-0002-3171-8889}$^\ddagger$, 
        Tianyi Li\,\orcidlink{0000-0001-5424-6442}$^\dagger$
    }
    \vspace{0.1cm} 

    \IEEEauthorblockA{
        $^\S$\textit{Shandong University, China} \quad
        $^\ddagger$\textit{Northeastern University, China} \quad
        $^\dagger$\textit{Aalborg University, Denmark}
    }
    \vspace{0.1cm}
    
    \IEEEauthorblockA{
        chenlinfang@mail.sdu.edu.cn, \{songzhen,l.liu,clz\}@sdu.edu.cn, \\
        \{guyu,Zhangyf,yuge\}@mail.neu.edu.cn, \{tianyi,yusli\}@cs.aau.dk
    }
    \vspace{-1cm}
}

\maketitle

\begin{abstract}
Memory-based Temporal Graph Neural Networks (M-TGNNs) maintain recursively updated node states to capture fine-grained temporal interactions. However, existing distributed frameworks lack effective mechanisms for managing the temporal data dependencies inherent in these models. As a result, they must enforce strict chronological updates, incur substantial remote synchronization overhead, and experience severe load imbalance when temporal event streams are skewed. We propose \texttt{DepTGL}, a scalable distributed training framework that restructures temporal-dependency management for M-TGNNs from a data-centric perspective. First, \texttt{DepTGL} introduces a hybrid temporal-dependency management scheme that explicitly balances communication and caching overhead via temporal-event caching, supplemented by selective dependency-driven communication. Next, \texttt{DepTGL} incorporates a gradient-aware cache-synchronization policy that adaptively suppresses boundary updates as model optimization stabilizes, thereby reducing redundant synchronization. Finally, \texttt{DepTGL} integrates a load-aware temporal-pruning strategy that eliminates auxiliary replay events under skew-induced load spikes, reducing redundant data processing and mitigating straggler effects. Experiments on six real-world temporal graphs show that \texttt{DepTGL} achieves an average speedup of 4.99$\times$ over state-of-the-art baselines, while maintaining comparable accuracy.
\end{abstract}


\section{Introduction}

Memory-based Temporal Graph Neural Networks (M-TGNNs) have achieved strong performance on dynamic graph learning tasks by directly modeling timestamped interaction events~\cite{nguyen2018continuous,kumar2019predicting,trivedi2019dyrep,xu2020inductive,rossi2020temporal}. 
Such timestamped continuous data streams are ubiquitous in various real-world applications, ranging from social network interactions to spatio-temporal mobility analysis, such as moving object clustering~\cite{evolutionary_clustering}, trajectory similarity computation~\cite{trajectory_similarity}, and transportation mode classification~\cite{estimator}.
Representative M-TGNN models, such as JODIE~\cite{kumar2019predicting}, DyRep~\cite{trivedi2019dyrep}, and TGN~\cite{rossi2020temporal}, maintain node-wise temporal states and update them along chronological interactions. 
Subsequent M-TGNN models enhance temporal representation learning by introducing mechanisms such as inductive temporal attention, asynchronous message propagation, and causal temporal walks, as exemplified by TGAT\cite{xu2020inductive}, APAN\cite{wang2021apan}, and CAW~\cite{wang2021inductive}.
While these models increase the expressive power of temporal graph representations, they also impose strict chronological dependencies across events and node states. Such dependencies hinder parallelization and complicate state management, making distributed training significantly more challenging than in static
GNN training~\cite{zheng2020distdgl,wan2022pipegcn,wang2022neutronstar,
lin2020pagraph,thorpe2021dorylus,wan2022bns,peng2023sancus}
or snapshot-based dynamic GNN training~\cite{song2024dynahb,swash,pgti}.

Distributed GNN systems, such as DistDGL~\cite{zheng2020distdgl}, PipeGCN~\cite{wan2022pipegcn}, NeutronStar~\cite{wang2022neutronstar}, PaGraph~\cite{lin2020pagraph}, Dorylus~\cite{thorpe2021dorylus}, BNS-GCN~\cite{wan2022bns}, and SANCUS~\cite{peng2023sancus}, have been extensively developed for large-scale static graphs. These systems focus on spatial dependencies and thus do not account for event ordering or evolving node states, making them unsuitable for memory-based TGNNs whose computation is tightly coupled to chronological interaction events. Recent distributed temporal-graph systems, including DynaHB~\cite{song2024dynahb}, SWASH~\cite{swash}, PGT-I~\cite{pgti}, and NeutronStream~\cite{chen2023neutronstream}, support dynamic workloads but operate on discrete-time snapshots or windowed updates rather than continuous event streams. As a result, they also fail to address the event-level temporal dependencies that fundamentally constrain the parallelization of memory-based TGNNs.

For event-driven M-TGNNs, distributed systems such as Sven~\cite{sven}, DistTGL~\cite{zhou2023disttgl}, DisTGL~\cite{fang2025distgl}, Cascade~\cite{cascade}, MemShare~\cite{memshare}, SWIFT~\cite{swift}, and TASER~\cite{taser} advance distributed temporal graph training through temporal partitioning, time-indexed storage, dependency-aware execution, shared state management, pipeline parallelism, and adaptive sampling. However, frequent data-state synchronization, rigid communication patterns, and hotspot-induced stragglers continue to form fundamental bottlenecks, limiting the scalability of distributed training for M-TGNNs on continuous-time dynamic graphs. We summarize the main design challenges in achieving parallel M-TGNN systems as follows.

\textbf{\textit{Challenge I: Cross-worker temporal data dependencies are difficult to serve efficiently.}}
In M-TGNNs, each node data state is recursively updated along its timestamped interaction history.
Processing a target event on one worker may therefore depend on earlier events and remote node states maintained by other workers.
Such dependencies break the sample independence assumed by
standard batch-parallel execution and make simple spatial caching
insufficient, because cached remote states cannot capture the missing chronological evolution between synchronization boundaries.
Handling these dependencies through online traversal or per-access remote fetching would place both dependency materialization and communication on the critical training path.

\textbf{\textit{Challenge II: Strict data consistency and cache synchronization incur massive communication overhead.}} Existing distributed TGNN frameworks typically enforce periodic or frequent cache-synchronization boundaries to maintain temporal data consistency across GPUs. This rigid cache synchronization mechanism not only slows down training significantly, but also ignores the convergence characteristics of the model in later stages, causing substantial avoidable data-state synchronization overhead and limiting the safe reuse of locally cached states.

\textbf{\textit{Challenge III: Data skewness in temporal streams leads to severe computational bottlenecks.}} Real-world temporal data streams typically exhibit highly skewed interaction patterns.
In distributed processing, both the input data volume and the dependency-induced data replay workload can fluctuate sharply across temporal batches.
Due to the reliance on global cache-synchronization barriers, fast workers must idle and wait for the heaviest-loaded worker, causing severe performance degradation and reduced cluster utilization.

Motivated by the above
challenges, we propose \texttt{DepTGL}, a parallel framework for distributed M-TGNN
training that combines lightweight remote caching,
in-window local replay, and adaptive boundary refreshes to reduce remote
data-state requests while preserving chronological update semantics. To address \textbf{\textit{Challenge I}}, \texttt{DepTGL} introduces
batch-wise hybrid temporal data dependency serving. It expands temporal data
dependencies offline, constructs worker-specific mixed batches, and performs
timestamp-ordered local replay, thereby moving dependency materialization
out of the runtime critical path. To address \textbf{\textit{Challenge II}},
\texttt{DepTGL} proposes gradient-aware dynamic cache synchronization. It
uses the EMA of gradient norms as a lightweight signal to skip stable boundary
refreshes and reduce redundant cross-worker synchronization. To address
\textbf{\textit{Challenge III}}, \texttt{DepTGL} designs load-aware temporal
data pruning. It uses the EMA of batch execution times to detect heavy-load
phases and prune auxiliary replay events before forward propagation while
preserving all target events. Our main contributions are summarized as
follows:
\begin{itemize}[leftmargin=*]
    \item We propose \texttt{DepTGL}, a parallel framework for distributed memory-based TGNN training. By combining offline temporal data dependency expansion, worker-specific mixed-batch construction, and timestamp-ordered local temporal data replay, \texttt{DepTGL} serves cross-worker temporal data dependencies outside the runtime critical path and reduces fine-grained remote data-state requests.
    \item We propose a gradient-aware dynamic cache synchronization mechanism that uses the moving average of gradient norms to adaptively control boundary synchronization, reducing unnecessary cache-synchronization operations as training stabilizes.
    \item We propose a load-aware temporal data pruning strategy that selectively
    prunes pure auxiliary replay events before forward propagation under
    heavy-load conditions, reducing prunable auxiliary computation and
    mitigating straggler effects at synchronization barriers.
    \item We implement \texttt{DepTGL} and evaluate it on real-world temporal
    graphs. Experimental results show that \texttt{DepTGL} achieves an average
    epoch-time speedup of 4.99$\times$ over state-of-the-art baselines, while
    maintaining comparable accuracy across datasets and models.
\end{itemize}
\section{Related Work}

\subsection{Temporal Graph Neural Networks}

Temporal graph neural networks can be broadly categorized into
Snapshot-based models discretize continuous graph evolution into a sequence of
static snapshots or temporal windows, typically capturing spatio-temporal
patterns by coupling graph convolutions with sequence models
~\cite{li2018diffusion,yu2018stgcn,zhao2020tgcn,zheng2020gman,pareja2020evolvegcn,sankar2020dysat,you2022roland}.
Capturing such spatio-temporal evolving patterns is also a fundamental challenge in related dynamic data mining fields, including evolutionary clustering of moving objects~\cite{evolutionary_clustering} and complex trajectory analyses~\cite{trajectory_similarity,estimator}.
From a computational perspective, the temporal dependencies in these methods
are coarse-grained and mainly bounded at the snapshot or window level.

In contrast, continuous-time methods process fine-grained timestamped
interaction events directly to capture structural and temporal dynamics
~\cite{nguyen2018continuous,xu2020inductive,wang2021apan,wang2021inductive}.
Among them, M-TGNNs, such as JODIE~\cite{kumar2019predicting},
DyRep~\cite{trivedi2019dyrep}, and TGN~\cite{rossi2020temporal},
maintain node-wise temporal data states and update them recursively along
timestamped interactions. Therefore, the data state required by a target
event depends on the chronological sequence of earlier events involving the
same nodes, which breaks the independence assumption of conventional batch-wise parallel training.

This distinction is important for distributed system design. Snapshot-based
systems optimize batching, partitioning, and caching around coarse-grained
temporal units, whereas M-TGNN training requires event-level chronological
data-state evolution. Therefore, efficient distributed M-TGNN training must
explicitly manage event ordering, remote temporal data-state access, and cache
consistency across workers during continuous event-stream processing.

\begin{figure*}[!t]
  \centering
  \includegraphics[width=0.86\textwidth]{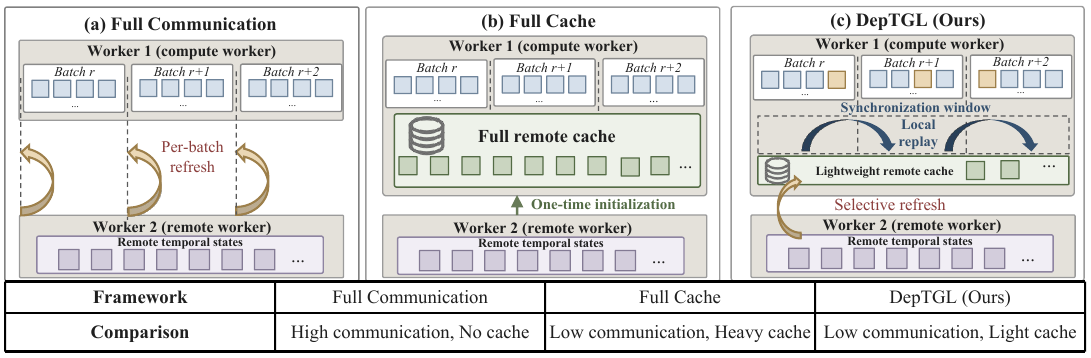}
  \caption{Motivating case study of communication-cache trade-offs.}
  \label{fig:case_study}
\end{figure*}
\vspace{1mm}

\subsection{Distributed Frameworks for Graph Learning}
Efforts to speed up graph training involve optimizations for single servers and distributed clusters across different graph representations. Table \ref{tab:comparison} summarizes representative graph training systems closely related to distributed and temporal graph learning.

\textbf{Static GNN training and system optimizations.}
Large-scale static GNN training has been extensively optimized through
neighbor sampling~\cite{hamilton2017inductive}, importance
sampling~\cite{chen2018fastgcn}, variance
reduction~\cite{chen2018stochastic}, graph
clustering~\cite{chiang2019cluster}, subgraph
sampling~\cite{zeng2020graphsaint}, and historical
embeddings~\cite{fey2021gnnautoscale}. These techniques build on canonical
message-passing GNN architectures such as GCN~\cite{kipf2017semi} and
GAT~\cite{velickovic2018graph}, and reduce the computation, sampling, or
memory overhead of static graph learning. System-level frameworks further improve scalability through distributed graph
storage~\cite{zhu2019aligraph,zheng2020distdgl,wang2021flexgraph}, feature
caching~\cite{lin2020pagraph}, pipelined feature
communication~\cite{wan2022pipegcn}, partition-parallel
training~\cite{wan2022bns}, boundary node
sampling~\cite{wan2022bns}, serverless execution~\cite{thorpe2021dorylus},
and staleness-aware communication avoidance~\cite{peng2023sancus}.
However, these systems mainly target static spatial dependencies, where node
representations do not maintain chronological memory states. Therefore,
their optimizations cannot be directly applied to memory-based TGNNs without
addressing event-order dependencies and temporal data-state consistency.

\textbf{Distributed training systems for discrete-time and spatiotemporal DGNNs.}
For discrete-time dynamic graphs, existing distributed DGNN systems mainly
optimize communication, batching, and partitioning around snapshot or
sliding-window execution. Early systems such as ESDG~\cite{chakaravarthy2021efficient},
DGC~\cite{chen2023dgc}, BLAD~\cite{fu2023blad}, and
DynaGraph~\cite{guan2022dynagraph} reduce communication or improve
parallelism through snapshot partitioning, chunk-based partitioning,
load-balanced scheduling, or sliding-window execution.
Recent systems further optimize this line of work: DynaHB~\cite{song2024dynahb}
uses hybrid batching for snapshot-based DGNNs, SWASH~\cite{swash}
reduces sliding-window communication through cache sharing, and
PGT-I~\cite{pgti} improves memory efficiency for spatiotemporal GNN
training.
These systems target snapshot-based or time-indexed graph execution, and
therefore differ from memory-based continuous-time TGNNs, where node data
states evolve along timestamped interaction events and must respect strict
chronological update semantics.

\textbf{Event-stream and continuous-time TGNN training systems.}
NeutronStream~\cite{chen2023neutronstream} supports event-stream dynamic
GNN training through sliding-window execution. DisTGL~\cite{fang2025distgl}
supports distributed temporal graph learning through streaming temporal-aware
edge partitioning, hybrid caching, and parallel fetching.
DistTGL~\cite{zhou2023disttgl} identifies the high overhead of distributed
data-state synchronization for memory-based TGNN training.
GNNFlow~\cite{gnnflow} improves continuous temporal GNN learning through
time-indexed graph storage, hybrid GPU-CPU placement, temporal sampling,
and dynamic caching.
Cascade~\cite{cascade} introduces temporal-data-dependency-aware execution,
while MemShare~\cite{memshare} mitigates communication bottlenecks through
hotspot data-state sharing.
Orthogonally, Sven and SWIFT accelerate TGNN execution through pipeline
parallelism on multi-GPU and single-machine architectures,
respectively~\cite{sven,swift}.
Additionally, TASER~\cite{taser} improves temporal adaptive sampling for
dynamic graph representation learning.
Despite these advancements, existing systems usually optimize only one or
two aspects, such as dependency scheduling, data-state sharing, sampling,
caching, or pipelining.
They still lack a unified runtime mechanism that jointly adapts cache
synchronization frequency and mitigates batch-level workload skew
for distributed memory-based TGNN training. More importantly, existing event-stream TGNN systems usually optimize
temporal dependency handling, cache management, and runtime scheduling
separately. In memory-based TGNNs, these factors are tightly coupled because
remote-state freshness, auxiliary replay, and worker-level load imbalance
jointly affect the execution cost of each batch.

From the perspective of remote temporal data-state serving, two intuitive
architectures are Full Communication and Full Cache. Full Communication keeps no persistent remote-state cache and relies on
frequent cross-worker communication to fetch or refresh remote temporal
states, while Full Cache reduces online communication by materializing remote
temporal histories locally. \texttt{DepTGL} takes a middle-ground
design that combines lightweight remote caching, selective boundary refreshes,
and local mixed-batch replay. Example~\ref{ex:case_study} illustrates the
communication-cache trade-off among the three designs.

\begin{example}[Communication--cache trade-off in remote temporal data-state serving]
\label{ex:case_study}
Figure~\ref{fig:case_study} compares three strategies for serving remote
temporal data states in distributed M-TGNN training. In
Figure~\ref{fig:case_study}(a), Full Communication keeps no persistent
remote-state cache and refreshes remote temporal states for each batch,
making cross-worker communication a dominant runtime cost. In
Figure~\ref{fig:case_study}(b), Full Cache materializes remote temporal
histories in advance, but the state required by a later batch, e.g., $B_T$,
may recursively depend on earlier events involving the same remote nodes
and trace back to $B_1$, causing large cache growth. In
Figure~\ref{fig:case_study}(c), \texttt{DepTGL} avoids
both extremes by dividing the event stream into synchronization windows of
$K$ batches, refreshing base remote states only at selected boundaries, and
reconstructing remote-state evolution through in-window local mixed-batch
replay.
\end{example}

\begin{table*}[t]
\centering
\caption{Comparison of representative temporal graph training systems.}
\label{tab:comparison}
\vspace{-1mm}

\small
\setlength{\tabcolsep}{2.6pt}
\renewcommand{\arraystretch}{1.08}

\begin{tabular*}{\textwidth}{@{\extracolsep{\fill}} l c c l c c c c @{}}
\toprule
\textbf{System} & \textbf{Target} & \textbf{Arch.} &
\textbf{Core Optimization} & \textbf{Comm.} & \textbf{Cache} &
\textbf{Adapt.} & \textbf{Load Bal.} \\
\midrule
DynaHB~\cite{song2024dynahb} & S-TGNN & Dist. &
Hybrid batching, vertex caching, async sync & Low & High & RL-based & Yes \\

SWASH~\cite{swash} & S-TGNN & Dist. &
Flexible comm., sliding-window cache sharing & Low & Med. & Adaptive sched. & Limited \\

PGT-I~\cite{pgti} & S-TGNN & Dist. &
Index-batching DDP & High & Low & No & Marginal \\

\midrule
SWIFT~\cite{swift} & M-TGNN & Single &
Data-state-aware pipeline & N/A & Med. & No & No \\

TASER~\cite{taser} & M-TGNN & Single &
Temporal adaptive sampling & N/A & Low & No & No \\

NeutronStream~\cite{chen2023neutronstream} & M-TGNN & Dist. &
Sliding-window event processing & Med. & Med. & No & Limited \\

DisTGL~\cite{fang2025distgl} & M-TGNN & Dist. &
STEP, hybrid cache, parallel fetch & Med. & Med. & Cache-aware & Limited \\

DistTGL~\cite{zhou2023disttgl} & M-TGNN & Dist. &
Data-state synchronization & High & Low & No & Limited \\

GNNFlow~\cite{gnnflow} & M-TGNN & Dist. &
Time-indexed storage, GPU-CPU cache & Med. & Med. & Cache-aware & Static \\

Cascade~\cite{cascade} & M-TGNN & Dist. &
Data-dependency-aware execution & Med. & Med. & Dep.-aware & Marginal \\

MemShare~\cite{memshare} & M-TGNN & Dist. &
Hotspot data-state sharing & Med. & Low & Static & Partial \\

Sven~\cite{sven} & M-TGNN & Dist. &
Load-balanced hierarchical pipeline & Med. & Low & No & Yes \\

\midrule
\textbf{DepTGL (Ours)} & \textbf{M-TGNN} & \textbf{Dist.} &
\textbf{Hybrid dep. serving, cache sync, pruning} &
\textbf{Low} & \textbf{Low} & \textbf{Yes} & \textbf{Yes} \\
\bottomrule
\end{tabular*}

\vspace{0.5mm}
\begin{minipage}{0.98\textwidth}
\footnotesize
\emph{Note:} S-TGNN denotes snapshot-based or spatiotemporal TGNNs, and
M-TGNN denotes memory-based or event-driven TGNNs. Comm. and Cache
qualitatively indicate residual communication pressure and cache/storage
footprint, respectively. Adapt. denotes runtime adaptivity. Limited/Partial
denotes static or heuristic implementations, and N/A denotes not applicable.
\end{minipage}
\vspace{-2mm}
\end{table*}

\subsection{Communication and Load Optimization}

Beyond temporal graph training frameworks, general distributed optimization
and graph learning systems provide useful techniques for communication
reduction and straggler mitigation in large-scale training.

\textbf{Communication Reduction.}
Communication reduction has been widely studied in distributed machine
learning through asynchronous updates~\cite{recht2011hogwild},
parameter-server training~\cite{dean2012large,li2014scaling}, local model
averaging~\cite{mcmahan2017communication,stich2018local}, and gradient
compression~\cite{lin2018deep}. 
In broader temporal and streaming data domains, data volume and system loads are frequently reduced through efficient time-series and trajectory compression techniques~\cite{camel,trace,uncertain_compression}. 
In distributed graph training, communication is reduced through graph partitioning~\cite{zheng2020distdgl},
feature caching~\cite{lin2020pagraph}, pipelined feature
communication~\cite{wan2022pipegcn}, boundary-node
sampling~\cite{wan2022bns}, and stale-cache reuse~\cite{peng2023sancus}.
For temporal graph training, recent systems further reduce communication
through sliding-window cache sharing~\cite{swash}, temporal-aware
partitioning and hybrid caching~\cite{fang2025distgl}, remote data-state
caching and time-indexed storage~\cite{zhou2023disttgl,gnnflow}, hotspot
data-state sharing~\cite{memshare}, dependency-aware
execution~\cite{cascade}, or pipeline-based execution~\cite{sven,swift}.
However, in memory-based TGNNs, communication reduction must also handle
chronological node data states and temporal messages whose validity depends
on timestamp-ordered updates. Existing systems reduce part of this cost
through caching, sharing, fetching, or pipelining, but mostly rely on fixed
or mechanism-specific policies and do not adapt temporal data-state
synchronization to runtime training dynamics across batches.

\textbf{Load Balancing and Straggler Mitigation.}
Load imbalance is another long-standing bottleneck in distributed graph
training. In static graph learning, workload skew is commonly mitigated
through graph clustering, subgraph sampling, boundary-node sampling, or
serverless execution~\cite{chiang2019cluster,zeng2020graphsaint,wan2022bns,thorpe2021dorylus}.
For dynamic and temporal graph training, DynaHB uses load-aware vertex
partitioning and load-balanced training for snapshot-based
DGNNs~\cite{song2024dynahb}; Sven accelerates distributed TGNN training
through load-balanced hierarchical pipeline parallelism~\cite{sven}; and
MemShare alleviates hotspot-related remote data-state communication through
hotspot data-state sharing~\cite{memshare}. From a broader distributed
optimization perspective, straggler-tolerant training also reduces the
computation assigned to slower workers to improve cluster
throughput~\cite{egger2023fast}. However, distributed memory-based TGNNs
have a more dynamic source of imbalance: each worker's load depends not only
on assigned target events, but also on auxiliary dependencies for remote-state
reconstruction. Such replay-induced workload can fluctuate sharply across
temporal batches, so runtime load control must distinguish supervised target
events from auxiliary replay events.
\section{Preliminaries}

\subsection{Temporal Data Stream and Node Data State}

We represent a temporal graph as a chronologically ordered temporal data
stream $\mathcal{G}=(\mathcal{V},\mathcal{E})$. Each event is denoted by
$e_k=(u_k,v_k,t_k,x_k)$, where $u_k$ and $v_k$ are the two endpoints,
$t_k$ is the timestamp, and $x_k$ is the event feature. Events are sorted
by time, i.e., $t_1 \le t_2 \le \cdots \le t_{|\mathcal{E}|}$.

Memory-based TGNNs, such as JODIE and TGN, maintain a memory vector
$m_u(t)$ for each node $u$, which serves as the node's temporal data state
and summarizes its temporal history. Given event $e_k$, the model
generates event-level messages and updates the memories of the involved
nodes recursively:



\vspace{-4mm}
\begin{align}
s_{u_k}^{(k)}, s_{v_k}^{(k)}
&= \mathrm{Message}(m_{u_k}(t_k^-),m_{v_k}(t_k^-),x_k,t_k),
\label{eq:message_generation}\\[0.5mm]
m_{u_k}(t_k)
&= \mathrm{Update}(m_{u_k}(t_k^-),s_{u_k}^{(k)}), \notag\\[0em]
m_{v_k}(t_k)
&= \mathrm{Update}(m_{v_k}(t_k^-),s_{v_k}^{(k)}).
\label{eq:memory_update}
\vspace{-2mm}
\end{align}

Here, $t_k^-$ denotes the time immediately before event $e_k$, and
$s_{u_k}^{(k)}$ and $s_{v_k}^{(k)}$ are the event messages used to update
the endpoint memories. The key implication is that a node representation depends on the ordered
sequence of its prior events, so different temporal batches may still be
coupled through temporal data states. This coupling creates cross-batch state
dependencies that must be preserved during distributed training to ensure correct temporal execution.

\subsection{Distributed Training Setting}

We consider a distributed training environment with $P$ workers.
The temporal graph nodes are disjointly partitioned across workers,
and each worker $p \in \{1,\dots,P\}$ owns a subset of nodes and
maintains their authoritative data states. During training, the
temporal data stream is divided into batches and processed
across the cluster. When an event involves a remote node, i.e., a node
owned by another worker, the processing worker must obtain the
corresponding historical data state of that node at the event
timestamp. 

To avoid requesting remote-node temporal data states from owner workers
whenever a remote node is accessed, distributed implementations often
maintain local cached copies of these states. However, these cached states may become stale as the
authoritative states continue to evolve on their owner workers. Therefore,
the processing worker must decide when and how to refresh its remote cache
so that the cached states remain temporally valid for subsequent events. A straightforward solution is to refresh remote data states frequently
through cross-worker cache synchronization. While this strategy can reduce
cache staleness and preserve temporal data consistency, it introduces
substantial communication traffic and synchronization overhead. 

\subsection{Problem Setup and Objective}
To illustrate how remote temporal data-state dependencies couple current
events with historical batches across workers, we provide
Example~\ref{ex:cross_worker_dependency}.

\vspace{-2mm}
\begin{figure}[H]
  \centering
  \includegraphics[width=\linewidth]{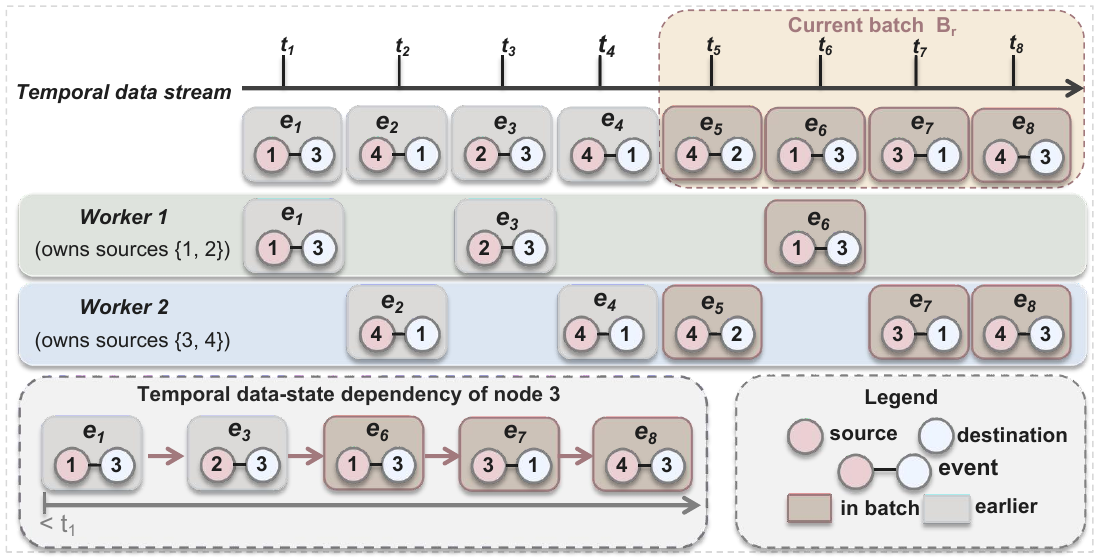}
  \vspace{-6mm}
  \caption{Example of cross-worker temporal data-state dependency.}
  \label{fig:temporal_coupling}
  \vspace{-2mm}
\end{figure}

\begin{example}[Cross-worker temporal data-state dependency]
\label{ex:cross_worker_dependency}
Figure~\ref{fig:temporal_coupling} illustrates why remote temporal
data-state caching is non-trivial in distributed M-TGNN training.
To process event $e_8=(4,3,t_8,x_8)$ in the current batch $B_r$, the
model needs the state of node~3 immediately before $t_8$. This state
depends on earlier updates to node~3, including $e_7$ and $e_6$ in
$B_r$ and $e_3$ and $e_1$ from previous batches. Since these events
may be assigned to different workers, the required state forms a
cross-worker and cross-batch dependency chain, making naive batch
parallelism and simple remote-state caching insufficient.
\end{example}

The dependency pattern illustrated above leads to three major system-level
bottlenecks. First, chronological node-data-state evolution induces temporal
data dependencies across current and historical batches, since the data
state required by an event may depend on a chain of earlier events involving
the same nodes. This limits parallelism and makes naive spatial data caching
insufficient. Second, preserving consistent remote data states often requires
frequent cross-worker cache synchronization, which can dominate the total
runtime. Third, skewed temporal data streams and bursty node interactions
create imbalanced workloads across workers, causing stragglers at
cache-synchronization barriers.

Our objective is to minimize the epoch training time of distributed
memory-based TGNNs while preserving the chronological node-data-state update
semantics required for model accuracy. Formally, the total execution
time of an epoch can be decomposed as
\vspace{-2mm}
\begin{equation}
T_{\mathrm{epoch}} =
\sum_{b}
\max_{p}
\left(
T_{\mathrm{comp}}^{p,b}
+
T_{\mathrm{comm}}^{p,b}
+
T_{\mathrm{wait}}^{p,b}
\right),
\vspace{-2mm}
\end{equation}
where $T_{\mathrm{comp}}^{p,b}$, $T_{\mathrm{comm}}^{p,b}$, and
$T_{\mathrm{wait}}^{p,b}$ denote the local computation time, the
cross-worker communication time, and the idle waiting time of worker
$p$ on batch $b$, respectively. \texttt{DepTGL} is designed to reduce these
three components jointly through temporal-data-dependency-aware execution and
adaptive runtime control.
\section{System Architecture and Hybrid Data Dependency Serving}
\label{sec:system}

\subsection{System Overview}

\begin{figure*}[t]
    \centering
    \includegraphics[width=0.72\textwidth]{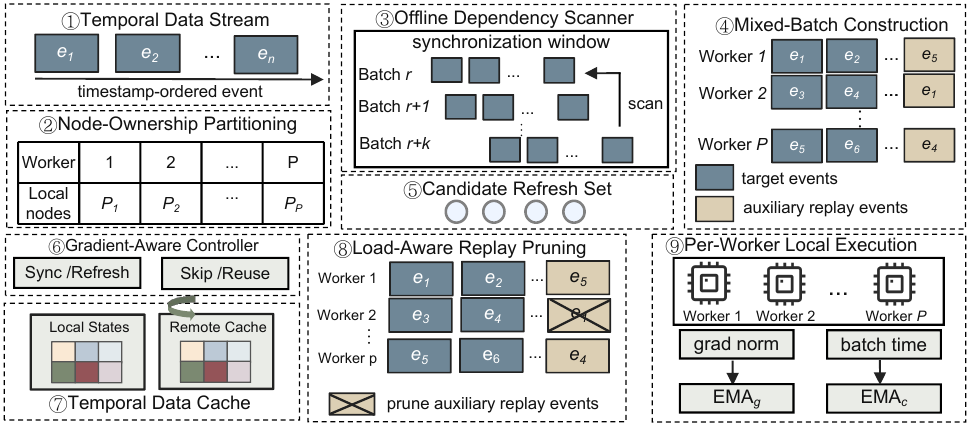}
    \vspace{-0.1cm}
    \caption{Overview of the DepTGL architecture.}
    \label{fig:architecture}
    \vspace{0mm}
\end{figure*}

Figure~\ref{fig:architecture} illustrates the overall workflow of \texttt{DepTGL}
according to the execution order of its main modules. Given a temporal graph
$\mathcal{G}=(\mathcal{V},\mathcal{E})$, \texttt{DepTGL} first organizes the input as a
timestamp-ordered temporal data stream (Step~\textcircled{1}) and partitions
nodes across workers based on node ownership (Step~\textcircled{2}). Each
worker maintains the authoritative temporal data states of its local nodes,
while remote-node states are served through local temporal data caches. Within
each synchronization window, the offline dependency scanner
(Step~\textcircled{3}) traces temporal dependencies backward from later target
events and identifies historical remote events required for local remote-state
reconstruction. \texttt{DepTGL} then constructs worker-specific mixed batches
(Step~\textcircled{4}), which contain target events for supervised training and
auxiliary replay events for advancing cached remote states, and generates
candidate refresh sets (Step~\textcircled{5}) that specify the remote nodes
whose base states may be refreshed at synchronization boundaries.

During runtime, \texttt{DepTGL} uses a gradient-aware controller
(Step~\textcircled{6}) to decide whether to refresh remote-node states or reuse
locally cached states. The temporal data cache (Step~\textcircled{7}) provides
local and cached remote states for timestamp-ordered execution. Before forward
propagation, load-aware replay pruning (Step~\textcircled{8}) removes eligible
pure auxiliary replay events under heavy-load conditions while preserving all
target events. Finally, each worker performs per-worker local execution
(Step~\textcircled{9}), computes the loss only over target events, and updates
model parameters together with runtime statistics used by subsequent adaptive
synchronization and pruning decisions.

\subsection{Offline Mixed-Batch Construction}
\label{sec:offline_construction}

Offline mixed-batch construction converts the in-window portion of remote
temporal data dependencies into worker-local auxiliary replay. For each
synchronization window, \texttt{DepTGL} identifies historical remote events, marks them
as auxiliary replay events, and combines them with target events to form
worker-specific mixed batches. Target events contribute to the supervised loss,
whereas auxiliary replay events only advance cached remote temporal data states.

\begin{definition}[Synchronization Window]
Given a synchronization interval $K$, a synchronization window
$\mathcal{W}_q$ is a consecutive group of temporal batches:
\vspace{-2mm}
\begin{equation}
\mathcal{W}_q =
\{B^{qK}, B^{qK+1}, \dots, B^{qK+K-1}\}.
\vspace{-2mm}
\end{equation}
The beginning of each window acts as a candidate cache-refresh boundary.
At this boundary, a worker may refresh the base states of required remote
nodes from their owner workers. Within the window, \texttt{DepTGL} advances temporal
data states through timestamp-ordered local mixed-batch execution.
\end{definition}

\begin{definition}[Mixed Batch]
For worker $p$ and batch index $r$, \texttt{DepTGL} constructs a mixed batch
as a timestamp-ordered execution sequence:
\vspace{-2mm}
\begin{equation}
\widetilde{B}_{p}^{r}
=
\operatorname{sort}_t
\left(
B_{p,\mathrm{tar}}^{r}
\cup
B_{p,\mathrm{aux}}^{r}
\right),
\label{eq:mixed_batch}
\vspace{-2mm}
\end{equation}
where $B_{p,\mathrm{tar}}^{r}$ denotes the target events assigned to
worker $p$, $B_{p,\mathrm{aux}}^{r}$ denotes the auxiliary replay events
identified by the offline dependency scan, and $\operatorname{sort}_t(\cdot)$
orders all events by timestamp before execution. Each event in
$\widetilde{B}_{p}^{r}$ is associated with an event-type mask that
distinguishes target events from auxiliary replay events.
\end{definition}

\begin{figure*}[!t]
    \centering
    \includegraphics[width=0.72\textwidth]{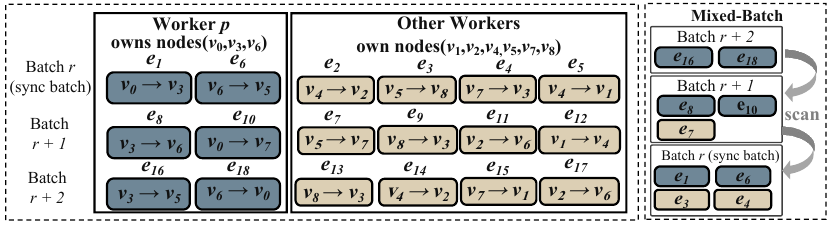}
    \vspace{-0.15cm}
    \caption{Example of backward dependency scanning and mixed-batch construction
    within a synchronization window. Blue boxes denote target events assigned to
    worker $p$, while beige boxes denote auxiliary replay events inserted for local
    remote-state reconstruction.}
    \label{fig:prefetcher}
    \vspace{0cm}
\end{figure*}

\begin{example}[Backward dependency scan]
Figure~\ref{fig:prefetcher} shows a synchronization window with three batches. For worker $p$, \texttt{DepTGL}
keeps its source-assigned target events and scans the window backward to
insert only the historical remote events needed for local remote-state
reconstruction. The resulting mixed batches preserve all target events and
add selected auxiliary replay events, enabling timestamp-ordered local
replay within the window.
\end{example}

Algorithm~\ref{alg:backward_scan} describes the offline construction of
worker-specific mixed batches. For each worker $p$, the algorithm identifies
the source-assigned target events that contribute to the supervised objective
and the auxiliary replay events needed to locally reconstruct remote temporal
data-state evolution within a synchronization window.

\begin{algorithm}[H]
\caption{Source-based Backward Dependency Scan and Mixed-Batch Construction}
\label{alg:backward_scan}
\begin{algorithmic}[1]
\REQUIRE Synchronization window $\mathcal{W}_q$; worker $p$;
global batches $\{B^s\}_{B^s\in\mathcal{W}_q}$;
node ownership mapping $\pi(\cdot)$
\ENSURE Mixed batches $\{\widetilde{B}_p^s\}_{B^s\in\mathcal{W}_q}$;
candidate boundary refresh set $\mathcal{N}_{p,\mathrm{sync}}^q$

\STATE $B_{p,\mathrm{tar}}^s \leftarrow \emptyset$,
$B_{p,\mathrm{aux}}^s \leftarrow \emptyset$ for each $B^s\in\mathcal{W}_q$
\STATE $\mathcal{F} \leftarrow \emptyset$;
\quad $\mathcal{I}_{\mathrm{aux}} \leftarrow \emptyset$

\FOR{each batch $B^s\in\mathcal{W}_q$ in reverse chronological order}
    \STATE $B_{p,\mathrm{tar}}^s \leftarrow
    \{e=(u,v,t,x)\in B^s \mid \pi(u)=p\}$
    \STATE $\mathcal{F} \leftarrow \mathcal{F} \cup
    \{v \mid e=(u,v,t,x)\in B_{p,\mathrm{tar}}^s,\ \pi(v)\ne p\}$

    \FOR{each event $e'=(u',v',t',x')\in B^s$}
        \IF{$(u'\in\mathcal{F}$ or $v'\in\mathcal{F})$ and $\pi(u')\ne p$}
            \IF{$\operatorname{id}(e')\notin\mathcal{I}_{\mathrm{aux}}$}
                \STATE Insert $e'$ into $B_{p,\mathrm{aux}}^s$
                \STATE $\mathcal{I}_{\mathrm{aux}} \leftarrow
                \mathcal{I}_{\mathrm{aux}}\cup\{\operatorname{id}(e')\}$
            \ENDIF
            \STATE $\mathcal{F} \leftarrow \mathcal{F}\cup
            \{i\in\{u',v'\}\mid \pi(i)\ne p\}$
        \ENDIF
    \ENDFOR
\ENDFOR

\STATE $\mathcal{N}_{p,\mathrm{sync}}^q \leftarrow \mathcal{F}$

\FOR{each batch $B^s\in\mathcal{W}_q$}
    \STATE $\widetilde{B}_p^s \leftarrow
    \operatorname{sort}_t(B_{p,\mathrm{tar}}^s\cup B_{p,\mathrm{aux}}^s)$
\ENDFOR

\RETURN $\{\widetilde{B}_p^s\}_{B^s\in\mathcal{W}_q}$,
$\mathcal{N}_{p,\mathrm{sync}}^q$
\end{algorithmic}
\end{algorithm}

The algorithm scans the batches in reverse chronological order. It first
collects the target events of worker $p$ and adds their remote destination
nodes to the dependency frontier $\mathcal{F}$ (Lines~4--5). Then, for each
event that touches the current frontier but is not a target event of worker
$p$, the algorithm inserts it as an auxiliary replay event and further expands
the frontier with its remote endpoints (Lines~6--12). This backward expansion captures historical events that may update the remote
states required by later target events. Inserted auxiliary replay events are marked by the
event-type mask and deduplicated during preprocessing, so they can be replayed
for temporal data-state reconstruction without being included in the
supervised loss.

After the scan, the remaining frontier forms the candidate boundary refresh
set $\mathcal{N}_{p,\mathrm{sync}}^q$ (Line~16). The target and auxiliary
events in each batch are then merged and sorted by timestamp to form the mixed
batch $\widetilde{B}_p^s$ (Lines~17--19). As a result, worker $p$ can execute
all supervised target events while replaying only the necessary historical
events for local temporal data-state reconstruction. This output further
provides the runtime controller with explicit refresh candidates and
target--auxiliary event masks.
\section{Adaptive Cache Synchronization and Load-aware Data Pruning}
\label{sec:adaptive_runtime}

The offline construction phase described in Section~\ref{sec:system}
moves most temporal data dependency expansion out of the runtime
critical path by constructing worker-specific mixed batches. However,
two runtime inefficiencies may still remain. First, even after offline
dependency expansion, candidate boundary cache refreshes may still
introduce non-negligible cross-worker communication and collective
waiting. Second, auxiliary replay events introduced by mixed-batch
construction may increase computation on hotspot batches, especially
under skewed temporal streams.

\texttt{DepTGL} addresses these two runtime issues through an adaptive runtime
controller. At candidate cache-refresh boundaries, \texttt{DepTGL} uses recent gradient
magnitude statistics to decide whether the current boundary refresh should
be executed or skipped. During local mixed-batch execution, \texttt{DepTGL} uses
recent batch-time statistics to detect hotspot-induced heavy load and
prunes eligible pure auxiliary replay events before they enter the
forward pass. Both decisions are deliberately conservative: target
events are never pruned, and skipped cache refreshes only reuse existing
local remote-cache states until the next executed refresh or the final
epoch-end synchronization.

\subsection{Gradient-aware Dynamic Cache Synchronization}
\label{subsec:dynamic_sync}

Boundary cache refreshes realign locally cached remote temporal data states
with their authoritative owner-side states. They are useful for reducing
remote-cache staleness, but executing them at every scheduled boundary can
be unnecessarily expensive. In memory-based TGNN training, early iterations
usually involve large parameter updates and rapidly changing node memories,
whereas later iterations tend to become more stable. When recent gradient
statistics indicate a stable update phase, immediately refreshing remote
data states at every boundary may bring limited benefit compared with its
communication and synchronization cost. \texttt{DepTGL} therefore uses gradient magnitude as a lightweight runtime signal to
adaptively reduce boundary refresh frequency during stable update phases.

Let $g_r$ denote the gradient norm computed after backpropagation on batch
$r$. In the distributed setting, this norm is computed from the synchronized
model gradients, so it provides a consistent training-dynamics signal for
the runtime controller. To capture recent training dynamics, \texttt{DepTGL}
maintains an exponential moving average (EMA) of gradient norms:
\vspace{-2mm}
\begin{equation}
  \bar{g}_r =
  \begin{cases}
    g_r, & \text{if } \bar{g}_{r-1}=0,\\
    \alpha g_r + (1-\alpha)\bar{g}_{r-1}, & \text{otherwise},
  \end{cases}
  \label{eq:grad_ema}
  \vspace{-2mm}
\end{equation}
where $\alpha$ is a smoothing factor. The EMA provides a stable estimate of
recent gradient dynamics and is used to evaluate upcoming candidate
cache-refresh boundaries.

For a scheduled candidate boundary at batch $r$, \texttt{DepTGL} makes the
synchronization decision using the most recently observed gradient norm
$g_{r-1}$ and its EMA $\bar{g}_{r-1}$:
\vspace{-1.5mm}
\begin{equation}
  S_r =
  \begin{cases}
    0, & r>0,\ \bar{g}_{r-1}>0,\ \text{and } g_{r-1}<\tau_g\bar{g}_{r-1},\\
    1, & \text{otherwise},
  \end{cases}
  \label{eq:sync_decision}
  \vspace{-2mm}
\end{equation}
where $S_r=1$ indicates that the boundary cache refresh is executed, and
worker $p$ fetches the needed remote-node data states in
$\mathcal{N}_{p,\mathrm{sync}}^q$ from their authoritative owner workers;
$S_r=0$ indicates that the refresh is skipped, and worker $p$ continues from
its existing local remote-cache states, relying on timestamp-ordered
mixed-batch execution to advance remote states locally within the
synchronization window. The threshold $\tau_g$ controls gradient stability:
when the recent gradient norm falls within a stable range relative to its EMA,
\texttt{DepTGL} skips the refresh; otherwise, it conservatively executes the refresh.
Since boundary cache refresh is a collective operation, the coordinator
broadcasts the synchronization decision as a constant-size Boolean flag to all
workers.

\subsection{Load-aware Temporal Data Pruning}

Offline mixed-batch construction inserts auxiliary replay events to reconstruct
remote temporal histories locally, but these events also increase local
computation and may amplify hotspot-induced stragglers. \texttt{DepTGL} mitigates this
issue through load-aware temporal data pruning: under heavy-load conditions,
it removes eligible pure auxiliary replay events before forward propagation
while preserving all target events.

Let $c_p^r$ denote the measured execution time of worker $p$ on batch $r$.
Since straggler behavior is worker-dependent, \texttt{DepTGL} maintains a worker-local
EMA of batch execution time:
\vspace{-2mm}
\begin{equation}
\bar{c}_p^r =
\begin{cases}
c_p^r, & \text{if } \bar{c}_p^{r-1}=0,\\
\beta c_p^r + (1-\beta)\bar{c}_p^{r-1}, & \text{otherwise},
\end{cases}
\label{eq:time_ema}
\vspace{-1.5mm}
\end{equation}
where $\beta$ is the smoothing factor. The EMA captures the recent execution
trend of each worker and provides a smoothed reference for detecting abnormal
local load.

Before executing batch $r$, \texttt{DepTGL} activates heavy-load mode when the latest
observed batch time exceeds the EMA-scaled baseline:
\vspace{-1.5mm}
\begin{equation}
H_p^r =
\begin{cases}
1, & r>0,\ \bar{c}_p^{r-1}>0,\ \text{and } c_p^{r-1}>\tau_c\bar{c}_p^{r-1},\\
0, & \text{otherwise},
\end{cases}
\label{eq:heavy_load}
\vspace{-1.5mm}
\end{equation}
where $H_p^r=1$ indicates that worker $p$ enters heavy-load mode for batch
$r$, and $\tau_c$ is a multiplicative load threshold.

Under the source-based assignment used by \texttt{DepTGL}, worker $p$ treats an event
$e=(u,v,t,x)$ as a target event when $\pi(u)=p$. Therefore, the eligible
pruning set of a mixed batch $\widetilde{B}_p^r$ is
\vspace{-2mm}
\begin{equation}
\mathcal{P}_p^r =
\{e=(u,v,t,x)\in \widetilde{B}_p^r \mid \pi(u)\ne p\}.
\label{eq:pruning_set}
\vspace{-2mm}
\end{equation}
These events correspond to pure auxiliary replay events for worker $p$ and do
not define its supervised target-event objective. When $H_p^r=1$, \texttt{DepTGL}
removes them before local forward propagation:
\vspace{-2mm}
\begin{equation}
\widehat{B}_p^r =
\begin{cases}
\widetilde{B}_p^r \setminus \mathcal{P}_p^r, & H_p^r=1,\\
\widetilde{B}_p^r, & H_p^r=0.
\end{cases}
\label{eq:pruned_batch}
\vspace{-2mm}
\end{equation}

The same source-based predicate is used to mask the training loss:
\vspace{-2mm}
\begin{equation}
M_p(e)=\mathbb{I}(\pi(u)=p), \quad e=(u,v,t,x).
\label{eq:target_mask}
\vspace{-2mm}
\end{equation}
The local batch loss is computed as
 \vspace{-1mm}
\begin{equation}
\mathcal{L}_p^r =
\begin{cases}
\frac{\sum_{e_i\in \widehat{B}_p^r} M_p(e_i)\ell(e_i)}
{\sum_{e_i\in \widehat{B}_p^r} M_p(e_i)},
& \text{if } \sum_{e_i\in \widehat{B}_p^r} M_p(e_i)>0,\\
0, & \text{otherwise},
\end{cases}
\label{eq:masked_loss}
\vspace{-1mm}
\end{equation}
where $\ell(e_i)$ is the edge prediction loss for event $e_i$.

Loss masking and pruning serve different purposes. Masking excludes auxiliary
replay events from the supervised objective, whereas pruning physically removes
eligible auxiliary replay events before forward propagation. Therefore,
pruning reduces auxiliary computation on heavy-load workers while preserving
the target-event training objective.

\subsection{Adaptive Runtime Controller}
\label{subsec:runtime_controller}

Algorithm~\ref{alg:adaptive_runtime} summarizes the runtime execution of
\texttt{DepTGL} over the mixed batches generated by
Algorithm~\ref{alg:backward_scan}. It adaptively controls two runtime
decisions: whether to refresh remote cached states at candidate boundaries,
and whether to prune auxiliary replay events under heavy local workload.

At each candidate refresh boundary, the coordinator invokes
$\mathrm{SyncDecision}$ using the previous gradient norm and its EMA
(Line~4). The decision is then broadcast to all workers (Line~5). If
synchronization is triggered, workers refresh the required remote states in
$\mathcal{N}_{p,\mathrm{sync}}^q$ from their owner workers (Lines~6--7);
otherwise, they continue using existing cached remote states. Since the
decision is broadcast as a Boolean flag, all workers follow the same
boundary-level synchronization policy.

For each batch, worker $p$ loads $\widetilde{B}_p^r$ and detects heavy load
using the previous batch time and its EMA (Lines~10--11). When heavy load is
detected, $\mathrm{PruneAux}$ removes eligible pure auxiliary replay events
before forward propagation (Line~12). Target events are always preserved.
The remaining events are executed in timestamp order, and the loss is
computed only on source-assigned target events (Lines~13--14). Finally, the
algorithm performs backpropagation, updates the gradient EMA and model
parameters, records the batch time, and updates the batch-time EMA for
subsequent adaptive decisions (Lines~15--19), followed by final temporal
data-state synchronization before evaluation (Line~21).

\begin{algorithm}[tbhp!]
\caption{Adaptive Runtime Controller of DepTGL}
\label{alg:adaptive_runtime}
\begin{algorithmic}[1]
\REQUIRE Mixed batches $\{\widetilde{B}_p^r\}$; candidate refresh sets
$\{\mathcal{N}_{p,\mathrm{sync}}^q\}$; node ownership map $\pi(\cdot)$;
thresholds $\tau_g,\tau_c$; EMA factors $\alpha,\beta$
\ENSURE Updated model parameters and temporal data states
\STATE Initialize $\bar{g}\leftarrow 0$, $g_{\mathrm{last}}\leftarrow 0$,
$\bar{c}_p\leftarrow 0$, $c_{p,\mathrm{prev}}\leftarrow 0$
\FOR{each batch $r$ in timestamp order}
  \IF{$r$ is a candidate cache-refresh boundary}
    \STATE $S_r \leftarrow \mathrm{SyncDecision}
    (g_{\mathrm{last}},\bar{g};\tau_g)$
    \STATE Broadcast $S_r$ from the coordinator to all workers
    \IF{$S_r=1$}
      \STATE Refresh remote cache entries in
      $\mathcal{N}_{p,\mathrm{sync}}^q$
    \ENDIF
  \ENDIF

  \STATE Load mixed batch $\widetilde{B}_p^r$
  \STATE $H_p^r \leftarrow \mathrm{LoadDetect}
  (c_{p,\mathrm{prev}},\bar{c}_p;\tau_c)$
  \STATE $\widehat{B}_p^r \leftarrow
  \mathrm{PruneAux}(\widetilde{B}_p^r,H_p^r,\pi)$
  \STATE Execute $\widehat{B}_p^r$ in timestamp order
  \STATE $\mathcal{L}_p^r \leftarrow
  \mathrm{MaskedLoss}(\widehat{B}_p^r,\pi)$
  \STATE Backpropagate $\mathcal{L}_p^r$ and compute gradient norm $g_r$
  \STATE $\bar{g}\leftarrow \mathrm{EMA}(\bar{g},g_r;\alpha)$;
  $g_{\mathrm{last}}\leftarrow g_r$
  \STATE Update model parameters
  \STATE Measure worker-local batch time $c_p^r$
  \STATE $\bar{c}_p\leftarrow \mathrm{EMA}(\bar{c}_p,c_p^r;\beta)$;
  $c_{p,\mathrm{prev}}\leftarrow c_p^r$
\ENDFOR
\STATE Synchronize final temporal data states before evaluation
\end{algorithmic}
\end{algorithm}

\section{THEORETICAL ANALYSIS}
\label{sec:theoretical_analysis}

\subsection{Cost and Straggler Analysis}
\label{subsec:cost_analysis}
\label{subsec:straggler_analysis}

The primary objective of \texttt{DepTGL} is to reduce the runtime of distributed
memory-based TGNN training by replacing fine-grained remote temporal
data-state requests with local mixed-batch execution, and by adaptively
reducing boundary communication and auxiliary replay computation.

\textbf{Communication Cost.}
Let $\mathcal{S}_K$ denote the set of scheduled cache-refresh boundaries under
synchronization interval $K$. For a boundary $r\in\mathcal{S}_K$, let $q(r)$
denote the corresponding synchronization window, and let
$\mathcal{N}_{p,\mathrm{sync}}^{q(r)}$ be the candidate remote-node set to be
refreshed by worker $p$. Let $d_m$ be the dimension of the temporal data state.
Under rigid boundary synchronization, worker $p$ incurs

\vspace{-2mm}
\begin{equation}
C_{\mathrm{rigid},p}
=
O\left(
d_m
\sum_{r\in\mathcal{S}_K}
\left|
\mathcal{N}_{p,\mathrm{sync}}^{q(r)}
\right|
\right).
\label{eq:rigid_comm_cost}
\end{equation}
\texttt{DepTGL} executes a boundary refresh only when the gradient-aware decision
$S_r=1$. Its boundary communication cost becomes
\begin{equation}
C_{\mathrm{DepTGL},p}
=
O\left(
d_m
\sum_{r\in\mathcal{S}_K}
S_r
\cdot
\left|
\mathcal{N}_{p,\mathrm{sync}}^{q(r)}
\right|
\right).
\label{eq:deptgl_comm_cost}
\end{equation}
\vspace{-1mm}
The communication ratio is
\vspace{-0.5mm}
\begin{equation}
\rho_{\mathrm{comm}}
=
\frac{
\sum_{r\in\mathcal{S}_K}
S_r
\cdot
\left|
\mathcal{N}_{p,\mathrm{sync}}^{q(r)}
\right|
}{
\sum_{r\in\mathcal{S}_K}
\left|
\mathcal{N}_{p,\mathrm{sync}}^{q(r)}
\right|
}
\le 1.
\label{eq:comm_ratio}
\end{equation}
\textbf{}
Thus, for the same candidate refresh sets, \texttt{DepTGL} never incurs more
boundary communication than rigid synchronization and reduces the effective
refresh frequency as more boundaries are skipped.

\textbf{Computation Cost.}
\texttt{DepTGL} trades part of the communication cost for local auxiliary replay.
Let $B_{p,\mathrm{tar}}^r$ and $B_{p,\mathrm{aux}}^r$ denote the target events
and auxiliary replay events in worker $p$'s mixed batch at batch $r$,
respectively. Without load-aware pruning, the local computation cost can be
approximated as
\vspace{-1mm}
\begin{equation}
T_{\mathrm{comp},p}^r
\approx
c_{\mathrm{tar}}
\left|
B_{p,\mathrm{tar}}^r
\right|
+
c_{\mathrm{aux}}
\left|
B_{p,\mathrm{aux}}^r
\right|,
\label{eq:comp_cost_before_pruning}
\end{equation}
where $c_{\mathrm{tar}}$ and $c_{\mathrm{aux}}$ are the average processing
costs of target and auxiliary replay events, respectively. With load-aware
pruning, eligible auxiliary replay events are removed when $H_p^r=1$, and the
computation cost becomes
\vspace{-1mm}
\begin{equation}
T_{\mathrm{comp},p}^{\prime r}
\approx
c_{\mathrm{tar}}
\left|
B_{p,\mathrm{tar}}^r
\right|
+
(1-H_p^r)c_{\mathrm{aux}}
\left|
B_{p,\mathrm{aux}}^r
\right|.
\label{eq:comp_cost_after_pruning}
\end{equation}
Therefore, load-aware pruning reduces auxiliary replay computation by
approximately
$H_p^r c_{\mathrm{aux}}\left|B_{p,\mathrm{aux}}^r\right|$, while preserving
all target events in the supervised objective.

\textbf{Straggler Effect.}
In synchronous distributed training, each batch is limited by the slowest
worker. Let the active execution time of worker $p$ on batch $r$ be
\vspace{-2mm}
\begin{equation}
A_p^r
=
T_{\mathrm{comp},p}^r
+
T_{\mathrm{comm},p}^r,
\label{eq:active_time}
\vspace{-2mm}
\end{equation}
and the corresponding iteration time be
\vspace{-2mm}
\begin{equation}
T_{\mathrm{iter}}^r
=
\max_p A_p^r.
\label{eq:iter_time}
\vspace{-2mm}
\end{equation}
The aggregate synchronization waiting cost is
\vspace{-2mm}
\begin{equation}
W_{\mathrm{wait}}^r
=
\sum_{p=1}^{P}
\left(
T_{\mathrm{iter}}^r - A_p^r
\right).
\label{eq:waiting_cost}
\vspace{-2mm}
\end{equation}

Temporal data skew may make some workers receive more auxiliary replay events
than others. When load-aware pruning is activated on worker $p$, the reduced
auxiliary computation can be approximated as
\vspace{-1.5mm}
\begin{equation}
\Delta_p^r
=
H_p^r c_{\mathrm{aux}}
\left|
B_{p,\mathrm{aux}}^r
\right|.
\label{eq:active_time_reduction}
\vspace{-1.5mm}
\end{equation}
The post-pruning active time is therefore
\vspace{-1.5mm}
\begin{equation}
A_p^{\prime r}
\approx
A_p^r - \Delta_p^r.
\label{eq:post_pruning_active_time}
\vspace{-1.5mm}
\end{equation}
If the pruned worker is a bottleneck or near-bottleneck worker, reducing
$A_p^r$ lowers $\max_p A_p^r$ or narrows the gap between workers, thereby
mitigating straggler-induced waiting.

\subsection{Controlled Approximation Analysis}
\label{subsec:approximation_analysis}

\texttt{DepTGL} introduces controlled approximation by skipping selected boundary cache
refreshes and pruning eligible pure auxiliary replay events. These decisions
may affect the freshness of locally cached remote temporal data states, but
they do not remove target events from the supervised training objective.

\textbf{Objective Preservation.}
\texttt{DepTGL} preserves the supervised target-event objective because load-aware
pruning removes only pure auxiliary replay events. According to the
source-based mask defined in Eq.~\eqref{eq:target_mask}, any pruned event
$e\in\mathcal{P}_p^r$ satisfies $M_p(e)=0$ and therefore does not contribute
to the local supervised loss in Eq.~\eqref{eq:masked_loss}. Thus, pruning
reduces auxiliary computation without removing target-event loss terms.

\textbf{Cache-State Approximation.}
Let $m_u(t)$ denote the authoritative temporal data state of remote node
$u$ at time $t$, and let $\widehat{m}_{p,u}(t)$ denote worker $p$'s locally
cached copy. The cache-state approximation error is
\vspace{-2mm}
\begin{equation}
\varepsilon_{p,u}(t)
=
\left\|
\widehat{m}_{p,u}(t)-m_u(t)
\right\|.
\label{eq:cache_error}
\vspace{-2mm}
\end{equation}
When a boundary refresh is executed and
$u\in\mathcal{N}_{p,\mathrm{sync}}^{q(r)}$, worker $p$ realigns the cached
remote state with the authoritative owner-side state:
\vspace{-2mm}
\begin{equation}
\widehat{m}_{p,u}(t_r)
\leftarrow
m_u(t_r),
\quad
u\in\mathcal{N}_{p,\mathrm{sync}}^{q(r)},\ S_r=1.
\label{eq:cache_realign}
\vspace{-2mm}
\end{equation}

Between two executed boundary refreshes, skipped refreshes and pruned
auxiliary replay events may make $\widehat{m}_{p,u}(t)$ deviate from the
authoritative state $m_u(t)$. However, this approximation affects only
remote-cache freshness rather than the set of supervised target events. Once a
later boundary refresh is executed, or when the final epoch-end synchronization
is performed before evaluation, the affected cached states are realigned with
their authoritative owner-side states.
\section{EXPERIMENTS}

\subsection{Experimental Setup}
\label{sec:experimental_setup}

\textbf{Datasets.}
We use six public real-world temporal graph datasets: AskUbuntu, MathOverflow, Reddit, MOOC, lastfm, and Wikipedia. These datasets cover diverse temporal interaction scenarios and differ in graph scale, event density, temporal duration, structural properties, and interaction sparsity. Table~\ref{tab:datasets} summarizes
their statistics. All datasets are publicly available from public sources.\footnote{\url{https://snap.stanford.edu}}
\vspace{-2mm}
\begin{table}[h]
\centering
\caption{Statistics of the evaluated temporal graph datasets.}
\label{tab:datasets}
\resizebox{1.0\columnwidth}{!}{
\begin{tabular}{lcccc}
\toprule
Dataset & Nodes & Events & Bipartite & Time Span \\
\midrule
AskUbuntu    & 159.3K & 964.4K & No  & 2613 days \\
MathOverflow & 24.8K  & 506.6K & No  & 2350 days \\
Reddit       & 11.0K  & 672.4K & Yes & 30 days \\
MOOC         & 7.0K   & 411.7K & Yes & 30 days \\
lastfm       & 2.0K   & 1.29M  & Yes & 1587 days \\
Wikipedia    & 9.2K   & 157.5K & Yes & 30 days \\
\bottomrule
\end{tabular}
}
\end{table}

\textbf{Models.}
We evaluate \texttt{DepTGL} with two representative memory-based continuous-time
TGNN models: TGN~\cite{rossi2020temporal} and
JODIE~\cite{kumar2019predicting}. TGN combines a node memory module with
temporal neighborhood aggregation, while JODIE updates user and item
embedding trajectories through recurrent interactions. Both models maintain
chronologically evolving node states, making them suitable for evaluating
distributed training under temporal data-state dependencies.

\textbf{Baselines.}
We compare \texttt{DepTGL} with four representative distributed TGNN
training baselines. For a controlled comparison, we implement their key
training policies within the same distributed runtime and use consistent
model settings across all methods. The implementation of \texttt{DepTGL} is
available in the public repository.\footnote{\url{https://github.com/LFchen111/DepTGL}}

\begin{itemize}[leftmargin=*]
  \item \textbf{Vanilla DDP.} Standard distributed data-parallel training
  with generic node partitioning and fixed-interval temporal memory
  synchronization. This baseline represents a common distributed execution
  mode used in temporal/dynamic graph training systems~\cite{fang2025distgl,
  chen2023neutronstream,memshare}.

  \item \textbf{DisTGL.} Uses STEP temporal partitioning and cache-based
  communication optimization for distributed temporal graph
  training~\cite{fang2025distgl}.

  \item \textbf{NeutronStream.} Uses sliding-window event processing to group
  temporally consecutive events in graph streams~\cite{chen2023neutronstream}.

  \item \textbf{MemShare.} Uses hotspot data-state sharing to reduce remote
  memory communication~\cite{memshare}.
\end{itemize}

\texttt{DepTGL} differs by jointly adapting mixed-batch dependency serving,
boundary cache synchronization, and auxiliary replay pruning at runtime.
For the memory--throughput analysis in Section~\ref{subsec:time_memory},
we additionally include a \textbf{Full Cache} variant as an analysis-only
reference that maximizes local materialization of remote temporal histories.

\textbf{Environments and Metrics.}
We conduct experiments in two hardware environments. Env~1 is the main real
distributed training environment, drawn from a resource pool of five
machines: four machines equipped with NVIDIA RTX 2080 Ti GPUs and one
machine equipped with an NVIDIA Tesla P100 GPU. Env~2 is a high-performance multi-GPU environment with 8 NVIDIA RTX 4090
GPUs.

We report epoch time, target-event throughput, speedup, test AUC, and test
AP. Throughput is computed as the number of original training events
divided by the average epoch time and does not count auxiliary replay
events as additional training progress. Speedup is computed over Vanilla
DDP using epoch time. For mechanism analysis, we further report
communication time, synchronization count, communication share, scaling
efficiency, and peak GPU memory.

\subsection{End-to-End Efficiency and Accuracy}
\label{subsec:end_to_end}

\begin{table*}[t]
\centering
\caption{End-to-end efficiency comparison.}
\label{tab:end2end_efficiency}

\begingroup
\scriptsize
\setlength{\tabcolsep}{1.2pt}
\renewcommand{\arraystretch}{0.84}
\def\tablethreescale{0.91}

\begin{minipage}[t]{0.492\textwidth}
\centering
\textbf{(a) Env~1: main real distributed environment.}

\vspace{0.4mm}

\begin{adjustbox}{scale=\tablethreescale,center}
\begin{tabular}{@{}ll*{6}{rr}@{}}
\toprule
\multirow{2}{*}{System} & \multirow{2}{*}{Model}
& \multicolumn{2}{c}{AskUb.}
& \multicolumn{2}{c}{MathOvf.}
& \multicolumn{2}{c}{Reddit}
& \multicolumn{2}{c}{MOOC}
& \multicolumn{2}{c}{lastfm}
& \multicolumn{2}{c}{Wiki} \\
\cmidrule(lr){3-4}
\cmidrule(lr){5-6}
\cmidrule(lr){7-8}
\cmidrule(lr){9-10}
\cmidrule(lr){11-12}
\cmidrule(lr){13-14}
& & ET & Spd. & ET & Spd. & ET & Spd. & ET & Spd. & ET & Spd. & ET & Spd. \\
\midrule
V-DDP & \multirow{5}{*}{JODIE}
& 668.41 & 1.00 & 492.22 & 1.00 & 270.90 & 1.00 & 805.60 & 1.00 & 855.21 & 1.00 & 380.08 & 1.00 \\
DisTGL
& & 655.02 & 1.02 & 491.15 & 1.00 & 214.28 & 1.26 & 651.06 & 1.24 & 680.03 & 1.26 & 311.92 & 1.22 \\
NS
& & 364.65 & 1.83 & 256.93 & 1.92 & 100.32 & 2.70 & 417.46 & 1.93 & 167.65 & 5.10 & 74.08 & 5.13 \\
MS
& & 612.24 & 1.09 & 371.24 & 1.33 & 134.69 & 2.01 & 597.78 & 1.35 & 231.91 & 3.69 & 100.46 & 3.78 \\
DepTGL
& & \textbf{47.54} & \textbf{14.06} & \textbf{93.84} & \textbf{5.25}
& \textbf{32.82} & \textbf{8.26} & \textbf{38.09} & \textbf{21.15}
& \textbf{87.40} & \textbf{9.79} & \textbf{20.43} & \textbf{18.61} \\
\midrule
V-DDP & \multirow{5}{*}{TGN}
& 737.96 & 1.00 & 542.55 & 1.00 & 339.28 & 1.00 & 887.88 & 1.00 & 987.28 & 1.00 & 408.58 & 1.00 \\
DisTGL
& & 720.63 & 1.02 & 537.99 & 1.01 & 271.58 & 1.25 & 727.44 & 1.22 & 767.36 & 1.29 & 343.39 & 1.19 \\
NS
& & 441.97 & 1.67 & 644.17 & 0.84 & 237.25 & 1.43 & 532.65 & 1.67 & 327.70 & 3.01 & 110.84 & 3.69 \\
MS
& & 254.35 & 2.90 & 422.65 & 1.28 & 185.85 & 1.83 & 681.14 & 1.30 & 362.04 & 2.73 & 133.13 & 3.07 \\
DepTGL
& & \textbf{119.62} & \textbf{6.17} & \textbf{90.20} & \textbf{6.02}
& \textbf{77.32} & \textbf{4.39} & \textbf{144.86} & \textbf{6.13}
& \textbf{160.18} & \textbf{6.16} & \textbf{55.83} & \textbf{7.32} \\
\bottomrule
\end{tabular}
\end{adjustbox}
\end{minipage}
\hfill
\begin{minipage}[t]{0.492\textwidth}
\centering
\textbf{(b) Env~2: high-performance multi-GPU environment.}

\vspace{0.4mm}

\begin{adjustbox}{scale=\tablethreescale,center}
\begin{tabular}{@{}ll*{6}{rr}@{}}
\toprule
\multirow{2}{*}{System} & \multirow{2}{*}{Model}
& \multicolumn{2}{c}{AskUb.}
& \multicolumn{2}{c}{MathOvf.}
& \multicolumn{2}{c}{Reddit}
& \multicolumn{2}{c}{MOOC}
& \multicolumn{2}{c}{lastfm}
& \multicolumn{2}{c}{Wiki} \\
\cmidrule(lr){3-4}
\cmidrule(lr){5-6}
\cmidrule(lr){7-8}
\cmidrule(lr){9-10}
\cmidrule(lr){11-12}
\cmidrule(lr){13-14}
& & ET & Spd. & ET & Spd. & ET & Spd. & ET & Spd. & ET & Spd. & ET & Spd. \\
\midrule
V-DDP & \multirow{5}{*}{JODIE}
& 68.92 & 1.00 & 27.60 & 1.00 & 70.88 & 1.00 & 41.54 & 1.00 & 45.40 & 1.00 & 18.50 & 1.00 \\
DisTGL
& & 68.39 & 1.01 & 26.57 & 1.04 & 58.30 & 1.22 & 33.72 & 1.23 & 34.05 & 1.33 & 15.86 & 1.17 \\
NS
& & 60.15 & 1.15 & 28.81 & 0.96 & 60.58 & 1.17 & 33.51 & 1.24 & 32.96 & 1.38 & 14.94 & 1.24 \\
MS
& & 60.85 & 1.13 & 30.60 & 0.90 & 57.82 & 1.23 & 33.60 & 1.24 & 30.38 & 1.49 & 14.72 & 1.26 \\
DepTGL
& & \textbf{52.47} & \textbf{1.31} & \textbf{16.36} & \textbf{1.69}
& \textbf{41.46} & \textbf{1.71} & \textbf{23.11} & \textbf{1.80}
& \textbf{24.19} & \textbf{1.88} & \textbf{10.49} & \textbf{1.76} \\
\midrule
V-DDP & \multirow{5}{*}{TGN}
& 127.73 & 1.00 & 77.61 & 1.00 & 91.47 & 1.00 & 125.00 & 1.00 & 153.10 & 1.00 & 52.42 & 1.00 \\
DisTGL
& & 124.55 & 1.03 & 75.96 & 1.02 & 83.04 & 1.10 & 117.74 & 1.06 & 140.08 & 1.09 & 49.11 & 1.07 \\
NS
& & 140.83 & 0.91 & 81.24 & 0.96 & 115.39 & 0.79 & 137.79 & 0.91 & 191.40 & 0.80 & 48.46 & 1.08 \\
MS
& & 125.30 & 1.02 & 73.04 & 1.06 & 86.99 & 1.05 & 126.43 & 0.99 & 161.65 & 0.95 & 45.26 & 1.16 \\
DepTGL
& & \textbf{107.36} & \textbf{1.19} & \textbf{64.41} & \textbf{1.21}
& \textbf{60.26} & \textbf{1.52} & \textbf{106.20} & \textbf{1.18}
& \textbf{128.41} & \textbf{1.19} & \textbf{43.45} & \textbf{1.21} \\
\bottomrule
\end{tabular}
\end{adjustbox}
\end{minipage}

\vspace{0.5mm}
\begin{minipage}{0.98\textwidth}
\footnotesize
\emph{Note:} ET denotes epoch time in seconds, and Spd. denotes speedup
over Vanilla DDP under the same dataset, model, and environment.
V-DDP, NS, MS, AskUb., MathOvf., and Wiki denote Vanilla DDP,
NeutronStream, MemShare, AskUbuntu, MathOverflow, and Wikipedia.
\end{minipage}

\endgroup
\vspace{-1.5mm}
\end{table*}

We first evaluate end-to-end efficiency and accuracy on temporal link
prediction. Table~\ref{tab:end2end_efficiency} reports epoch time and
speedup over Vanilla DDP in Env~1 and Env~2, while
Table~\ref{tab:end2end_accuracy} reports new-node test AUC/AP under the
same settings. For each dataset and model, the comparison uses the four
baselines described in Section~\ref{sec:experimental_setup}.

Table~\ref{tab:end2end_efficiency}(a) shows that, in the main distributed
environment, \texttt{DepTGL} achieves the lowest epoch time among all
compared systems across all datasets and both memory-based TGNN models. Under the
default TGN setting, \texttt{DepTGL} obtains 4.39--7.32$\times$ speedups over
Vanilla DDP, with an average speedup of 6.03$\times$. \texttt{DepTGL} also
outperforms MemShare by up to 4.70$\times$ on MOOC and 4.69$\times$ on
MathOverflow. These results suggest that hotspot data-state sharing
alleviates remote memory communication, but does not fully address the
combined overhead of frequent cache refreshes, temporal dependency serving,
and straggler-induced waiting. In contrast, \texttt{DepTGL} combines
dependency-aware mixed-batch replay, gradient-aware boundary
synchronization, and load-aware auxiliary replay pruning to reduce
fine-grained remote data-state requests, boundary communication, and
straggler-induced stalls.

The JODIE results further show that the benefit is not tied to a single
TGNN architecture. Since both JODIE and TGN maintain recursively updated
node memories, these results indicate that \texttt{DepTGL}'s adaptive temporal data
dependency management is broadly effective for memory-based
continuous-time TGNNs.

Table~\ref{tab:end2end_efficiency}(b) reports the results under Env~2. The absolute speedups are smaller
because Env~2 provides higher GPU throughput and lower communication
latency. Nevertheless, \texttt{DepTGL} still achieves the best epoch time on every
dataset, demonstrating that the proposed mechanisms remain effective beyond
communication-constrained environments.

\begin{table}[H]
\centering
\scriptsize
\setlength{\tabcolsep}{1.7pt}
\renewcommand{\arraystretch}{0.92}
\caption{New-node link prediction accuracy comparison.}
\label{tab:end2end_accuracy}
\begin{adjustbox}{width=0.98\columnwidth,center}
\begin{tabular}{@{}lccccc@{}}
\toprule
Dataset & V-DDP & DisTGL & NS & MS & DepTGL \\
\midrule
\multicolumn{6}{c}{\textbf{Env~1, JODIE}} \\
\midrule
AskUbuntu
& .7144/.7506 & \textbf{.7182/.7539} & .7125/.7493 & .7111/.7465 & .7153/.7501 \\
MathOverflow
& .7189/.7390 & .7199/.7381 & \textbf{.7224/.7425} & .7094/.7297 & .7199/.7390 \\
Reddit
& .8783/.8791 & .8775/.8746 & \textbf{.8817/.8853} & .8746/.8743 & .8815/.8819 \\
MOOC
& .6591/.6265 & \textbf{.6620/.6330} & .6566/.6139 & .6523/.6242 & .6578/.6231 \\
lastfm
& .7628/.7477 & .7893/.7872 & \textbf{.7932/.7989} & .7827/.7847 & .7863/.7864 \\
Wikipedia
& .7543/.7471 & .8005/.8019 & \textbf{.8042/.8170} & .7813/.7862 & .7982/.8049 \\
\midrule
\multicolumn{6}{c}{\textbf{Env~1, TGN}} \\
\midrule
AskUbuntu
& .7390/.7944 & .7598/.8078 & .7622/.8070 & \textbf{.7691/.8044} & .7548/.7997 \\
MathOverflow
& .7228/.7475 & .7482/.7723 & \textbf{.7594/.7861} & .7214/.7495 & .7402/.7552 \\
Reddit
& .9384/.9409 & .9495/.9532 & \textbf{.9497/.9519} & .9317/.9278 & .9418/.9461 \\
MOOC
& .7865/.7700 & \textbf{.8422/.8168} & .7491/.7008 & .8271/.8086 & .8268/.8158 \\
lastfm
& .7151/.7198 & .7810/.7805 & .7865/.7990 & .7869/.8038 & \textbf{.7908/.8047} \\
Wikipedia
& .9276/.9367 & .9329/.9409 & \textbf{.9493/.9559} & .9312/.9402 & .9410/.9484 \\
\midrule
\multicolumn{6}{c}{\textbf{Env~2, JODIE}} \\
\midrule
AskUbuntu
& .7067/.7423 & \textbf{.7152/.7510} & .7090/.7454 & .7081/.7402 & .7140/.7478 \\
MathOverflow
& .7171/.7375 & .7101/.7265 & \textbf{.7186/.7393} & .6812/.7010 & .7145/.7321 \\
Reddit
& .8557/.8423 & .8596/.8416 & .8712/.8636 & .8307/.8066 & \textbf{.8716/.8657} \\
MOOC
& .6432/.6144 & .6520/.6188 & \textbf{.6523/.6078} & .6359/.6064 & .6375/.5984 \\
lastfm
& .7489/.7331 & .7888/.7865 & \textbf{.7921/.7967} & .7717/.7746 & .7847/.7837 \\
Wikipedia
& .7353/.7184 & .7886/.7948 & \textbf{.7969/.8062} & .7674/.7616 & .7933/.8031 \\
\midrule
\multicolumn{6}{c}{\textbf{Env~2, TGN}} \\
\midrule
AskUbuntu
& .7480/.7935 & .7839/.8269 & \textbf{.7856/.8218} & .7587/.8004 & .7568/.8076 \\
MathOverflow
& .7313/.7477 & .6741/.7195 & .7436/.7641 & .7199/.7440 & \textbf{.7612/.7903} \\
Reddit
& .9060/.9097 & .9399/.9440 & .9189/.9254 & .9365/.9406 & \textbf{.9457/.9494} \\
MOOC
& .7023/.6535 & .8644/.8446 & .8350/.8186 & .8446/.8229 & \textbf{.8813/.8679} \\
lastfm
& .7596/.7780 & .7630/.7588 & .6960/.6997 & .7605/.7720 & \textbf{.8096/.8244} \\
Wikipedia
& .9028/.9142 & .9418/.9483 & \textbf{.9457/.9529} & .9369/.9450 & .9392/.9471 \\
\bottomrule
\end{tabular}
\end{adjustbox}
\vspace{1mm}
\begin{minipage}{0.98\columnwidth}
\footnotesize
\vspace{1mm}
\emph{Note:} Each entry reports AUC/AP. V-DDP, NS, and MS denote
Vanilla DDP, NeutronStream, and MemShare.
\end{minipage}
\vspace{-2mm}
\end{table}

Table~\ref{tab:end2end_accuracy} shows that the efficiency gains of \texttt{DepTGL}
do not come at the cost of systematic accuracy degradation. In the default
TGN setting under Env~1, \texttt{DepTGL} improves over Vanilla DDP on all six
datasets in both AUC and AP. For example, on lastfm, \texttt{DepTGL} increases
AUC/AP from 0.7151/0.7198 to 0.7908/0.8047. On MOOC, it improves AUC/AP
from 0.7865/0.7700 to 0.8268/0.8158.

This behavior is consistent with \texttt{DepTGL}'s design. \texttt{DepTGL} always preserves
target events for supervised loss computation, while skipped boundary
refreshes only reuse locally reconstructed remote states and load-aware
pruning only removes pure auxiliary replay events. Therefore, the proposed
runtime optimizations reduce synchronization and replay overhead without
changing the target-event learning objective.

\subsection{Time Breakdown and Memory Use}
\label{subsec:time_memory}

\begin{figure*}[t]
  \centering
  \includegraphics[
    width=0.82\textwidth,
    trim={0 0.55cm 0 0},
    clip
  ]{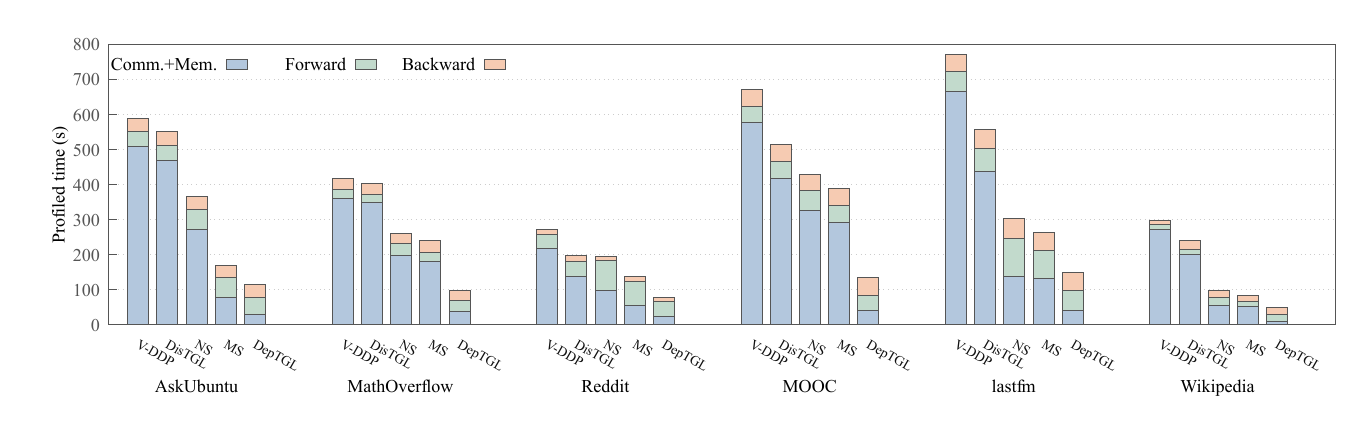}
  \vspace{-1mm}
  \caption{Profiled-time breakdown across datasets.}
  \label{fig:time_breakdown}
  \vspace{-1mm}
\end{figure*}

We analyze runtime composition under Env~1 with TGN. Table~\ref{tab:time_breakdown}
reports a profiled-time breakdown on AskUbuntu, and
Figure~\ref{fig:time_breakdown} gives the cross-dataset view. We profile
communication, runtime batch preparation, forward propagation, backward
propagation, and memory management, together with synchronization count and
communication share.

\vspace{-2mm}
\begin{table}[H]
\centering
\caption{Representative profiled-time breakdown.}
\label{tab:time_breakdown}
\footnotesize
\setlength{\tabcolsep}{2.0pt}
\renewcommand{\arraystretch}{1.05}
\begin{tabular*}{\columnwidth}{@{\extracolsep{\fill}}lcccccccc@{}}
\toprule
System & \#Sync & Comm. & Prep. & Fwd. & Bwd. & Mem. & Total & Comm.\% \\
\midrule
Vanilla DDP   & 3215.0 & 487.64 & 0.25 & \textbf{42.21} & 37.90 & 20.09 & 588.09 & 82.9 \\
DisTGL        & 3215.0 & 445.09 & \textbf{0.24} & 42.24 & 40.52 & 24.63 & 552.73 & 80.5 \\
NeutronStream & 1608.0 & 247.51 & 0.31 & 56.91 & 38.45 & 23.36 & 366.54 & 67.5 \\
MemShare      & 322.0  & 56.57  & 0.32 & 57.42 & \textbf{36.25} & \textbf{19.74} & 170.30 & 33.2 \\
DepTGL        & \textbf{26.1} & \textbf{7.57} & 0.56 & 48.35 & 37.46 & 21.50 & \textbf{115.45} & \textbf{6.6} \\
\bottomrule
\end{tabular*}

\vspace{0.5mm}
\begin{minipage}{\columnwidth}
\footnotesize
\emph{Note:} Times are in seconds. \#Sync reports the average synchronization count per epoch. Comm.\% is computed as Comm./Total.
\end{minipage}
\vspace{-2mm}
\end{table}

Table~\ref{tab:time_breakdown} shows that communication dominates Vanilla DDP
and DisTGL on AskUbuntu: both perform 3215.0 synchronizations per epoch and
spend over 80\% of the profiled time on communication. NeutronStream and
MemShare reduce part of this cost, but their communication shares remain
67.5\% and 33.2\%. In contrast, \texttt{DepTGL} reduces the synchronization
count to 26.1 and communication time to 7.57s, cutting total runtime from
588.09s to 115.45s and reducing the communication share from 82.9\% to 6.6\%.

Figure~\ref{fig:time_breakdown} provides a cross-dataset view. For readability,
Comm.+Mem. combines cross-worker memory synchronization and memory-management
time, while runtime batch preparation is omitted because it is negligible.
Across all datasets, \texttt{DepTGL} substantially reduces the dominant
communication and memory-related state-management cost.

\vspace{-3mm}
\begin{table}[H]
\centering
\caption{Memory--throughput trade-off.}
\label{tab:memory_throughput}
\vspace{-0.3em}
\small
\setlength{\tabcolsep}{8pt}
\renewcommand{\arraystretch}{1.08}
\begin{tabular}{lcccc}
\toprule
System & ET & Thr. & Spd. & Mem. \\
       & (s) & (events/s) & (x) & (GB) \\
\midrule
Vanilla DDP    & 1107.23 & 466  & 1.00 & 2.44 \\
DisTGL         & 1070.53 & 482  & 1.03 & 2.56 \\
NeutronStream  & 363.31  & 1419 & 3.05 & 2.96 \\
MemShare       & 382.40  & 1348 & 2.90 & 2.58 \\
Full Cache     & 254.13  & 2029 & 4.36 & 3.10 \\
\textbf{DepTGL} & \textbf{176.51} & \textbf{2921} & \textbf{6.27} & \textbf{2.44} \\
\bottomrule
\end{tabular}
\vspace{-0.5mm}
\begin{flushleft}
\footnotesize
\emph{Note:} ET denotes epoch time; Thr. denotes target-event throughput;
Spd. denotes speedup over Vanilla DDP; Mem. denotes peak GPU memory.
\end{flushleft}
\vspace{-3mm}
\end{table}

Table~\ref{tab:memory_throughput} quantifies the communication-cache trade-off
on AskUbuntu, with Full Cache as an analysis-only reference. \texttt{DepTGL}
achieves the highest throughput, reaching 2921 events/s, 6.27$\times$ higher
than Vanilla DDP and 2.17$\times$ higher than MemShare, while using the same
peak GPU memory as Vanilla DDP. Full Cache increases peak memory to 3.10GB
but still achieves lower throughput than \texttt{DepTGL}, showing that
maximizing local materialization is less effective than balancing local
replay, boundary refresh control, and auxiliary replay pruning.

\vspace{-2mm}
\subsection{Ablation Study and Hyperparameter Sensitivity }
\label{subsec:ablation}

\begin{figure}[H]
    \centering
    \includegraphics[width=\columnwidth]{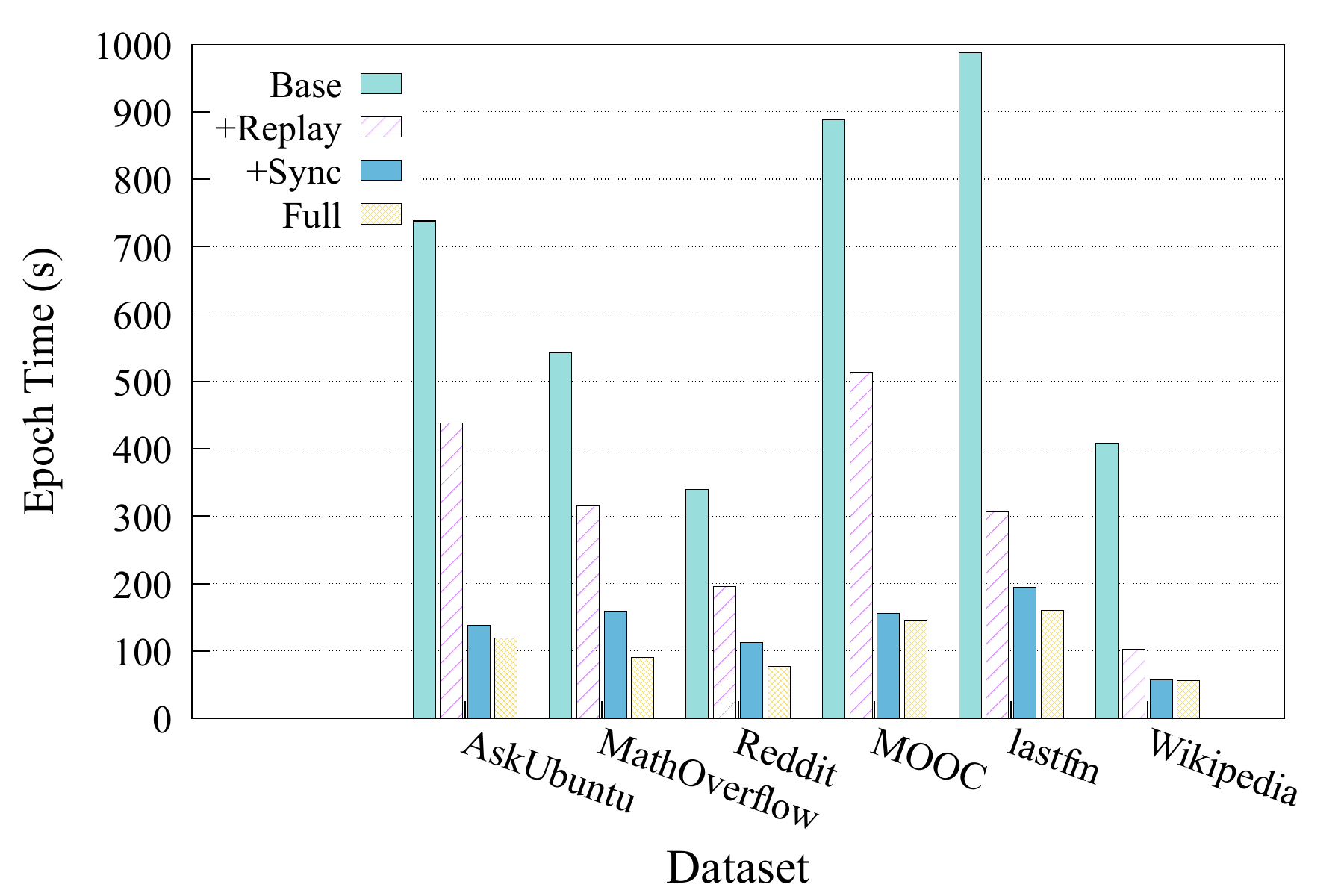}
    \caption{Ablation study.}
    \label{fig:ablation}
    \vspace{-2mm}
\end{figure}

We conduct ablation experiments to isolate the contribution of each \texttt{DepTGL}
component. Figure~\ref{fig:ablation} reports four incremental
configurations under the main distributed environment with TGN:
``Base'' disables all \texttt{DepTGL} mechanisms; ``+Replay'' enables
offline mixed-batch construction and worker-local temporal replay with
fixed synchronization; ``+Sync'' further enables gradient-aware
dynamic cache synchronization; and ``Full'' additionally enables
load-aware temporal data pruning to form the complete \texttt{DepTGL}
configuration.

As shown in Figure~\ref{fig:ablation}, the three components provide
complementary benefits. First, ``+Replay'' moves temporal dependency expansion
into offline mixed-batch construction and serves many remote temporal
data-state dependencies through local replay. On Wikipedia, it reduces epoch
time from 408.58s to 103.02s, achieving a 3.97$\times$ speedup over ``Base''.
Second, ``+Sync'' adapts boundary cache-refresh decisions to recent gradient
dynamics. On AskUbuntu, it reduces epoch time from 437.85s to 137.81s,
showing that fixed synchronization remains costly even after local replay is
enabled. Third, ``Full'' prunes eligible pure auxiliary replay events under
heavy-load conditions, further reducing Reddit epoch time from 112.62s to
77.32s over ``+Sync''. Overall, the three components jointly reduce remote
data-state access, boundary communication, and batch-level workload skew.

\vspace{-4mm}

We further conduct a compact hyperparameter sensitivity study on
MathOverflow and lastfm. Table~\ref{tab:hyperparameter_test} reports the results when
varying one hyperparameter at a time around the reference configuration
$K=6$, $\tau_g=1.6$, and $\tau_c=0.5$. Here, $K$ is the synchronization
window length, $\tau_g$ controls boundary-refresh skipping, and $\tau_c$
controls load-aware pruning under batch-level runtime skew.

\vspace{-0.4em}
\begin{table}[H]
\centering
\caption{Hyperparameter sensitivity.}
\label{tab:hyperparameter_test}
\normalsize
\setlength{\tabcolsep}{3.8pt}
\renewcommand{\arraystretch}{0.92}
\begin{tabular*}{\columnwidth}{@{\extracolsep{\fill}}lccc rr rr}
\toprule
\multirow{2}{*}{Cfg.}
& \multirow{2}{*}{$K$}
& \multirow{2}{*}{$\tau_g$}
& \multirow{2}{*}{$\tau_c$}
& \multicolumn{2}{c}{MathOverflow}
& \multicolumn{2}{c}{lastfm} \\
\cmidrule(lr){5-6}\cmidrule(lr){7-8}
& & & & Time & Slow. & Time & Slow. \\
\midrule
\#1  & 3  & 1.6 & 0.5 & 70.806 & 8.5\%  & 138.754 & 10.3\% \\
\#2  & 20 & 1.6 & 0.5 & 75.332 & 15.5\% & 129.424 & 2.9\%  \\
\#3  & 6  & 1.0 & 0.5 & 73.958 & 13.4\% & 133.952 & 6.5\%  \\
\#4  & 6  & 2.5 & 0.5 & 76.410 & 17.1\% & 142.404 & 13.2\% \\
\#5  & 6  & 1.6 & 0.0 & 71.912 & 10.2\% & 131.608 & 4.6\%  \\
\#6  & 6  & 1.6 & 1.0 & 74.620 & 14.4\% & 139.752 & 11.1\% \\
\midrule
Ref. & 6  & 1.6 & 0.5 & \textbf{65.240} & 0.0\% & \textbf{125.790} & 0.0\% \\
\bottomrule
\end{tabular*}
\vspace{1mm}

\begin{minipage}{0.98\columnwidth}
\footnotesize
\textit{Note:} Time is average epoch time; Slow. denotes slowdown relative
to the reference configuration.
\end{minipage}
\vspace{-2mm}
\end{table}

The results show that moderate runtime-control parameters provide the best
balance. A too-small $K$ introduces more refresh opportunities, while a
too-large $K$ may increase auxiliary replay. Overly conservative or aggressive
$\tau_g$ weakens the balance between communication saving and cache-state
freshness, and extreme $\tau_c$ either prunes too aggressively or leaves
runtime skew insufficiently mitigated.

\subsection{Evaluation of Scalability}
\label{subsec:scalability}

We evaluate scalability by varying the number of workers under Env~1 with
TGN. We report target-event throughput and scaling efficiency, computed as
$\mathrm{Eff}_n=\mathrm{Thr}_n/(n\cdot \mathrm{Thr}_1)$.

\vspace{-2mm}
\begin{table}[H]
\centering
\caption{Scalability comparison.}
\label{tab:scalability}
\scriptsize
\setlength{\tabcolsep}{2.0pt}
\renewcommand{\arraystretch}{0.96}
\begin{adjustbox}{width=0.98\columnwidth,center}
\begin{tabular}{@{}lrrrrr@{}}
\toprule
System & 1W Thr. & 2W Thr. & 3W Thr. & 2W Eff. & 3W Eff. \\
\midrule
\multicolumn{6}{c}{\textbf{AskUbuntu}} \\
\midrule
Vanilla DDP
& \textbf{2341.24} & 1113.57 & 464.34  & 23.78\% & 6.61\%  \\
DisTGL
& 2269.94 & 1111.95 & 480.57  & 24.49\% & 7.06\%  \\
NeutronStream
& 2311.03 & 1845.86 & 1419.54 & 39.94\% & 20.47\% \\
MemShare
& 2321.56 & 1790.70 & 1288.33 & 38.57\% & 18.50\% \\
DepTGL
& 2283.51 & \textbf{2838.78} & \textbf{2882.40}
& \textbf{62.16\%} & \textbf{42.08\%} \\
\midrule
\multicolumn{6}{c}{\textbf{Reddit}} \\
\midrule
Vanilla DDP
& 2389.08 & 1257.90 & 503.87  & 26.33\% & 7.03\%  \\
DisTGL
& 2360.56 & 1397.19 & 623.23  & 29.59\% & 8.80\%  \\
NeutronStream
& 2305.24 & 1949.06 & 1164.91 & 42.27\% & 16.84\% \\
MemShare
& \textbf{2409.96} & 1994.78 & 1356.34 & 41.39\% & 18.76\% \\
DepTGL
& 2382.33 & \textbf{3424.28} & \textbf{2928.71}
& \textbf{71.87\%} & \textbf{40.98\%} \\
\bottomrule
\end{tabular}
\end{adjustbox}
\vspace{0.5mm}

\begin{minipage}{0.98\columnwidth}
\footnotesize
\textit{Note:} The one-worker setting is
used as the reference for computing scaling efficiency within each system. 1W, 2W, and 3W denote one, two, and three workers, respectively.
\end{minipage}
\vspace{-2mm}
\end{table}

Table~\ref{tab:scalability} shows that \texttt{DepTGL} maintains the strongest
multi-worker scaling behavior. With two workers, \texttt{DepTGL} achieves
2838.78 target events/s on AskUbuntu and 3424.28 target events/s on Reddit,
corresponding to 62.16\% and 71.87\% scaling efficiency, respectively.
With three workers, \texttt{DepTGL} further reaches 2882.40 target events/s on
AskUbuntu and 2928.71 target events/s on Reddit, outperforming Vanilla DDP by
6.21$\times$ and 5.81$\times$, and MemShare by 2.24$\times$ and 2.16$\times$,
respectively. Notably, only \texttt{DepTGL} improves from two to three workers
on AskUbuntu, suggesting better preservation of useful parallelism.

The baselines exhibit different scalability bottlenecks. Vanilla DDP and
DisTGL are dominated by frequent temporal data-state synchronization, while
NeutronStream and MemShare reduce part of synchronization or hotspot-state
communication but still lag behind \texttt{DepTGL}. This suggests that
existing baselines alleviate individual overheads, but remain less effective
when remote temporal data-state access, boundary refreshes, and batch-level
workload skew jointly determine multi-worker M-TGNN execution cost.

\section{Conclusion}
\label{sec:conclusion}

We presented \texttt{DepTGL}, a parallel framework for
distributed memory-based TGNN training with adaptive temporal data dependency
management. By combining batch-wise hybrid temporal data dependency serving,
gradient-aware dynamic cache synchronization, and load-aware temporal data
pruning, \texttt{DepTGL} reduces remote temporal data-state access,
cache-refresh communication, and skew-induced execution overhead. Experiments
on six real-world temporal graphs show that \texttt{DepTGL} achieves an
average speedup of 4.99$\times$ over state-of-the-art baselines while
maintaining comparable accuracy.

\section*{AI-Generated Content Acknowledgement}
AI-based tools were used for language polishing and figure draft refinement.
All technical content was reviewed and verified by the authors.

\bibliographystyle{IEEEtran}
\bibliography{IEEEabrv,references}

\end{document}